\documentstyle[epsfig,lscape]{aipproc}

\def\etal{{et\,al.}}
\def\msun{M$_{\odot}$}

\def\mdot{$\dot M$}
\def\grad{$^\circ$}

\def\gro{GRO J1655-40}
\def\grs{GRS 1915+105}

\begin{document}

\title{Microquasars}

\author{Jochen Greiner}
\address{Astrophysical Institute
        Potsdam, An der Sternwarte 16, 14482 Potsdam, Germany}

\maketitle

\begin{abstract}
Microquasars are binary systems in our Galaxy which sporadically eject
matter at relativistic speeds along bipolar jets. While phenomenologically 
similar to quasars, their vicinity allows much more in-depth studies.
I review the observational aspects of microquasars,
in particular the new developments on the interplay between
accretion disk instabilities and jet ejection.
\end{abstract}

\section*{1. Introduction}

Accreting X-ray binaries  can be classified
[1] in various ways, two important being those according to the nature of the
compact object (neutron star or black hole) or the mass of the donor
(low-mass vs. high-mass). Most of the high-mass X-ray binaries 
(HMXB) show X-ray pulsations thus suggesting a magnetized neutron star
as the accretor. The donors usually are giants of spectral typ O or B.
Low-mass X-ray binaries (LMXB) comprise X-ray bursters, globular cluster
X-ray sources, soft X-ray transients and many galactic bulge sources.
The donors typically are main-sequence stars of spectral types K-M.
Both, neutron stars as well as black hole candidates have been found as
accreting objects in LMXBs.

Radio observations with high spatial resolution during the past decade 
have shown that a small number of X-ray binaries display blobs which 
move with an apparent velocity larger than the velocity of light
(superluminal motion) away from the core of the X-ray source 
(Fig. \ref{radio1655}). 
Though the lifetime of these blobs as radio emitters is short (few days to
weeks) compared to the repetition timescale of ejections, the generally 
accepted notion is that two-sided jets are emitted by these X-ray sources.
The phenomenological similarity to radio-loud quasars led to the naming
of microquasars [2].

There is no strict limit above what jet speed a system is considered 
to be a microquasar. Tab. 1 lists all galactic binaries with 
jets faster than 0.1c (though often the speed is not exactly known). 
As can be seen, microquasars do not belong to one of the above X-ray source
categories, since they comprise both, either neutron star or black hole
accretors as well as low-mass or high-mass donors.

Previous reviews with different emphasis on individual sources,
jet physics and observational constraints can be found in [3, 4].

\begin{landscape}
\begin{table}
 \caption{Galactic binary sources showing relativistic jets ($v> 0.1$c)}
  \begin{tabular}{lccccccccc}
\noalign{\smallskip}
Source & X-ray$^{(1)}$ & Radio$^{(1)}$ & Accretor & Donor & D (kpc) & 
   $V_{app}$$^{(2)}$ & $V_{int}$$^{(3)}$ &   $\Theta^{(4)}$ & Refs. \\
\noalign{\smallskip}
\hline
\noalign{\smallskip}
GRS 1915+105  & t & t & black hole?    &           & 10--12 & 
     1.2c-1.7c & 0.92c-0.98c & 66$^{\circ}$-70$^{\circ}$ & [5, 6] \\
GRO J1655-40 (V1033 Sco) & t & t & black hole    & F3-6 IV   & 3.2    & 
     1.1c & 0.92c & 72$^{\circ}$-85$^{\circ}$ & [7, 8, 9] \\
XTE J1748-288 & t & t & black hole?    &           & 8--10  & 
     0.9c-1.5c & $>$0.93c  &  & [10] \\
XTE J1819--254 (V4641 Sgr)$^{(5)}$    & t & t &               &     & 0.5?   & 
     $\sim$0.8c &      &   & [11] \\
SS 433 (V1343 Aql)    & t & t & neutron star & OB?       & 5.5    & 
     0.26c & 0.26c & 79$^{\circ}$ & [12, 13] \\
Cygnus X-3 (V1521 Cyg)   & t & t & neutron star? & WR?       & 8--10  & 
     $\sim$0.3c & $\sim$0.3c & $>$70$^{\circ}$ & [14, 15] \\
CI Cam (XTE  J0421+560)       & t & t & neutron star? & Be?       &$\sim$1 & 
     $\sim$0.15c & $\sim$0.15c  & $>$70$^{\circ}$ & [16, 17] \\
Circinus X-1 (BR Cir) & p & p & neutron star  & MS        & 8.5    & 
     $\geq$0.1c & $\geq$0.1c & $>$70$^{\circ}$ & [15, 18] \\
1E1740.7-2942 & p & p & black hole?    &           & 8--10  & 
                & & & [2, 19] \\
GRS 1758-258  & p & p & black hole?    &           & 8--10  & 
       & & & [20, 21] \\
\noalign{\smallskip}
\hline
\end{tabular}

\noindent{\small

 $^{(1)}$  t $\equiv$ transient, p $\equiv$ persistent \\
 $^{(2)}$  V$_{app}$ is the apparent speed of the highest velocity component 
           of the ejecta. \\
 $^{(3)}$  V$_{int}$ is the intrinsic velocity of the ejecta. \\
 $^{(4)}$  $\Theta$ is the angle between the direction of motion 
            of the ejecta and the line of sight. \\
     $^{(5)}$ This X-ray transient has initially been related to the optical
              variable GM Sgr. However, GM Sgr is a different variable, and
              the optical counterpart of XTE J1819--254 was named 
              V\,4641 Sgr [22].
    }
\label{muquas}

\end{table}
\end{landscape}

\section*{2. The two most important microquasars}

Our present understanding mostly rests on the results obtained over the
last years for two microquasars, namely \grs\ and \gro. Therefore, some
basics concerning these two systems will be described below.

\subsection*{2.1. \grs}

\begin{figure}[hb!]
\vspace{-0.3cm}
\begin{minipage}[c]{0.7\textwidth}
\centerline{\epsfig{file=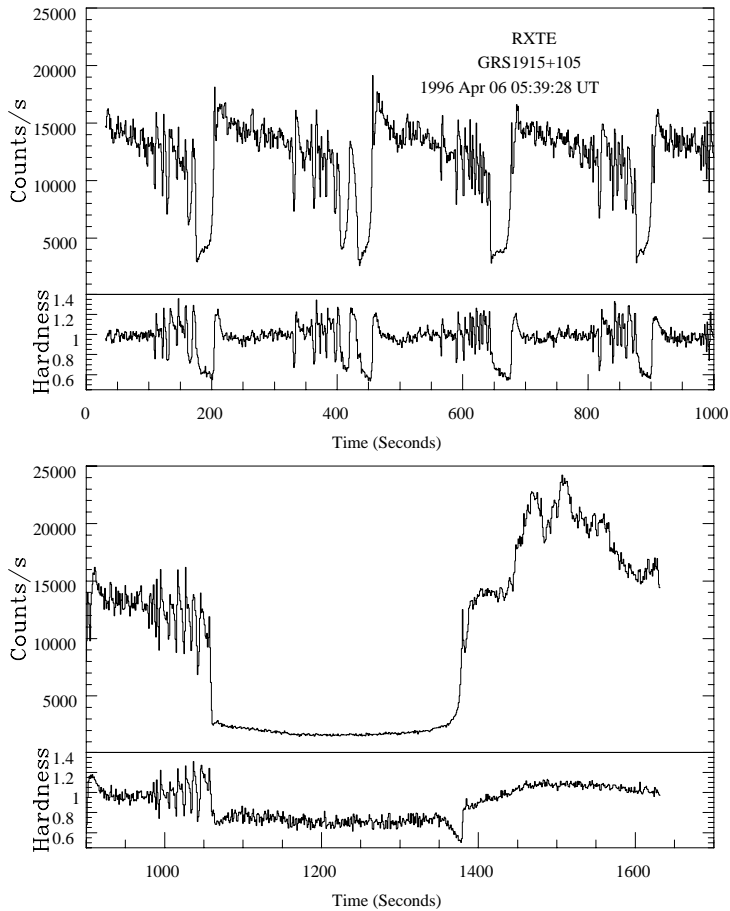,width=.96\textwidth}}
\end{minipage}%
\begin{minipage}[c]{0.3\textwidth}
\vspace{3.5cm}
\caption{X-ray light curve of GRS 1915+105 
        on April 6, 1996 showing the quasi-periodically repeating pattern of 
        30 sec duration brightness sputters (top) and a major lull (bottom).
       The top panel of each plot shows the count rate while
       the lower panel shows the hardness ratio (count ratio in the 4.4-25 keV
       versus 2-4.4 keV band) at the same time resolution of 1 sec
       (time along the abscissa refers to the time labeled in the top panel).
       (Taken from [25].)
 }
\label{stall}
\end{minipage}
\end{figure}

Already shortly after the discovery of \grs\ as a strongly variable new
X-ray transient in 1992 [23] it became clear that it was an unusual source.
It did not show the canonical fast-rise, exponential decay light curve,
and did not show a related bright optical outburst. Later observations
showed it to be X-ray active even during times of BATSE-non-detections,
suggesting a strongly variable X-ray spectrum [24]. With the
RXTE observations since early 1996 these spectral and temporal variations
were discovered [25] to extend down to timescales of seconds to minutes 
(Fig. \ref{stall}),
leading to a variety of new interpretations (see par. 3--5), including the
emptying of the inner part of the accretion disk during the stalls
and its subsequent replenishment [25, 26].

High-resolution radio observations in 1994 revealed for the first time
superluminal motion of plasmons in our Galaxy [5]. Recent observations [6]
revealed plasmon velocities substantially larger than those observed in 1994,
and constrain the distance to less than about 11.5 kpc. Since the distance
of \grs\ is not accurately known, the constraint on the bulk Lorentz factor
is very loose (1.4--30). Due to its location in the galactic plane and a
consequently huge visual extinction ($\sim$25--27 mag) no decent optical 
studies are possible. Thus, nothing is known yet about the orbital period and
the binary components.

\subsection*{2.2. \gro}

\begin{figure}[ht!]
\vspace{-.3cm}
\begin{minipage}[c]{0.65\textwidth}
\centerline{\epsfig{file=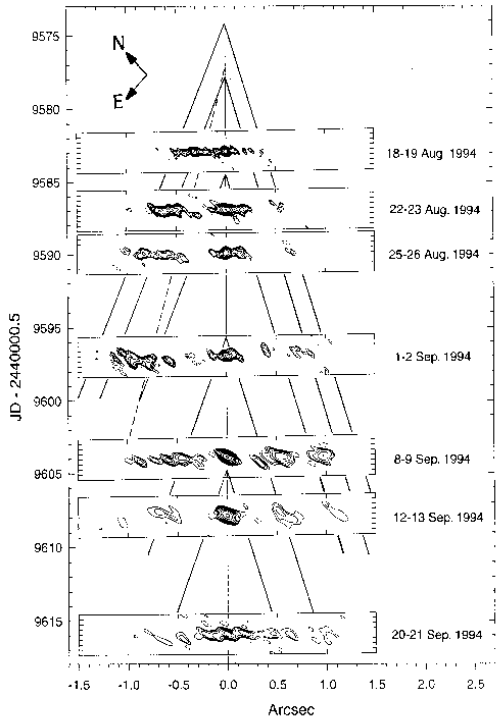,width=0.86\textwidth}}
\end{minipage}%
\begin{minipage}[c]{0.3\textwidth}
\vspace{3.cm}
\caption{A sequence of seven VLBA images of \gro\ at 1.6 GHz, each having
   an agular size of 3\farcs0$\times$0\farcs4. The vertical separation 
   corresponds to the time elapsed between the images. The solid lines
   between the images identify motions of 54 mas/d (left) and 
   45.5 mas/d (right). The vertical line marks the central source from which
   the offset (in arcsec) is shown on the horizontal axis (from [8]).} 
\label{radio1655}
\end{minipage}
\end{figure}

\gro\ was discovered in July 1994 as a new X-ray transient [27], and 
superluminal motion of the radio blobs (Fig. \ref{radio1655}) was discovered 
within a few weeks [7, 8]. However, the interpretation of the radio data
is much more complicated as compared to \grs, since the blob movement
shows wiggles of about 2\grad\ around the jet axis, and the relative 
brightness of the receding and approaching jet does not follow the simple
relativistic Doppler boosting description. Instead, the observed brightness
difference is often much larger, and also flips from side to side [8].
Thus, the ejecta must be intrinsically asymmetric. Furthermore, the 
jet axis seems not to be perpendicular to the orbital plane, but inclined
by 15\grad\ [9].

The  brightness ($\sim$17 mag) of the optical counterpart has allowed to 
determine with high precision [9] the orbital period (2\fd621),
the inclination (69\fdg5), the mass (2.34 \msun) and radial velocity amplitude 
(228.2 km/s) of the donor, and finally the mass of the accreting object
of 7.02$\pm$0.22 \msun.  With this mass estimate, the accretor in \gro\
is one of the best black hole candidates in the Galaxy.

\section*{3. Quasi-periodic oscillations}

RXTE observations have revealed a perplexing variety of  quasi-periodic
oscillations (QPO) in the X-ray power density spectra
[28, 29, 30]. Some sources show up to 7 different QPOs simultaneously
spanning three decades in frequency; QPOs
sometimes occur  with up to 3 harmonics, and their power has
a diverse energy dependence.
While QPOs are believed to provide a valuable means of probing
the X-ray emission region, the understanding of basic properties is still
rather poor. 

Often, QPOs are attributed to processes in the accretion disk,
but evidence for such an origin is scarce. In several cases the properties
of QPOs seem to correlate much better with those of hard X-rays, commonly
believed to arise in a Comptonizing corona, rather than those of soft X-rays
(accretion disk). An intriguing correlation
has recently been found in \grs\ [31]: during source states with QPOs
the X-ray intensity variations are mostly in the hard, power law component,
while during states without QPOs intensity variations are dominated by
the soft, accretion disk component. In addition, there is a strong
correlation between QPO frequency (2--10 Hz) and disk temperature.
This suggests a delicate interplay between accretion disk and 
Comptonizing corona in a way where the QPO is produced in the corona,
but its frequency is determined by the state of the disk.

One of the simple predictions based on the existence of a Comptonizing corona
is the time lag of hard X-ray photons with respect to soft ones.
This is indeed observed [32]. Moreover, the observed time lag scales roughly 
logarithmically with photon energy, as would be expected from a
Comptonization process. However, a detailed QPO waveform analysis
for 4 QPOs in \grs\ has shown that the mean waveform does not exhibit the 
profile smearing that would be caused at the delayed higher energies [28].

A ``stable'' QPO has been seen in both, \grs\ (67 Hz) and \gro\ (300 Hz) 
with varying strength. If associated with the Keplerian motion at the
last stable orbit around a (non-rotating) black hole according to
$f$ (kHz)$ = 2.2 / M_{\rm BH}$ (\msun) gives $M_{\rm BH}\sim7$ \msun\
for \gro, in surprising agreement with the optically determined mass!
However, the strong increase of the fractional rms towards larger energies
(up to 25 keV) is incompatible with an origin in the disk which has an
effective temperature of 2 keV. Alternatively, three other models
have been proposed, all resorting on relativistic effects (e.g. [32] for
an overview): (i) diskoseismic oscillations [33, 34],
(ii) frame dragging [35] and (iii) oscillations related to a 
centrifugal barrier [36].
At present, it is not clear which of these models, if any, provides
a correct description.

\section*{4. Rotating black holes?}

The X-ray spectra of microquasars consist of (at least) four different
components [37]:
(1) a thermal component with effective temperature of 2--3 keV which 
usually is attributed to the emission of the accretion disk,
(2) a hard, power law component extending up to 600 keV [38] 
without any
obvious cut-off which is generally interpreted as comptonization
of the accretion disk spectrum by hot electrons in a corona above the
disk,
(3) iron features which have been interpreted as absorption lines of
 He- and H-like iron [39, 40], and 
(4) an additional component comprising of
excess emission in the 10--20 keV range which has been interpreted as
Compton reflection hump.

The effective temperature of the thermal  component is very high when
compared to the neutron star binaries ($\approx$1.2 keV) or canonical 
black hole soft X-ray transients ($\approx$0.7--1.1 keV).
If interpreted according to the standard accretion disk prescription
of Shakura \& Sunyaev [41], where T (keV) = 1.2 (\mdot/M)$^{1/4}$,
the observed temperatures of 2--3 keV would correspond to a mass of
the central object of much less than 1 \msun. Such a low mass,
however, seems unacceptable given the high luminosities and non-thermal
spectra up to 600 keV. Thus, either a different cooling process is
active in microquasars, or the implicit assumption of the last stable
orbit at 3 Schwarzschild radii for the above $T(M)$ relation is not valid.
In fact, for prograde rotating black holes the innermost stable orbit
is closer to the hole, thus allowing higher temperatures of the disk.
Application to \gro\ with its known parameters yields a nearly maximally
rotating black hole (a=0.93) [42]. For \grs, assuming a mass of
30 \msun\ (based on the lower QPO frequency and luminosity arguments),
a similar high rotation rate is deduced (a=0.998). Note that this 
black hole spin
is not inconsistent with the Thirring-Lense interpretation of the QPOs.

\begin{figure}[th!]
\centerline{\epsfig{file=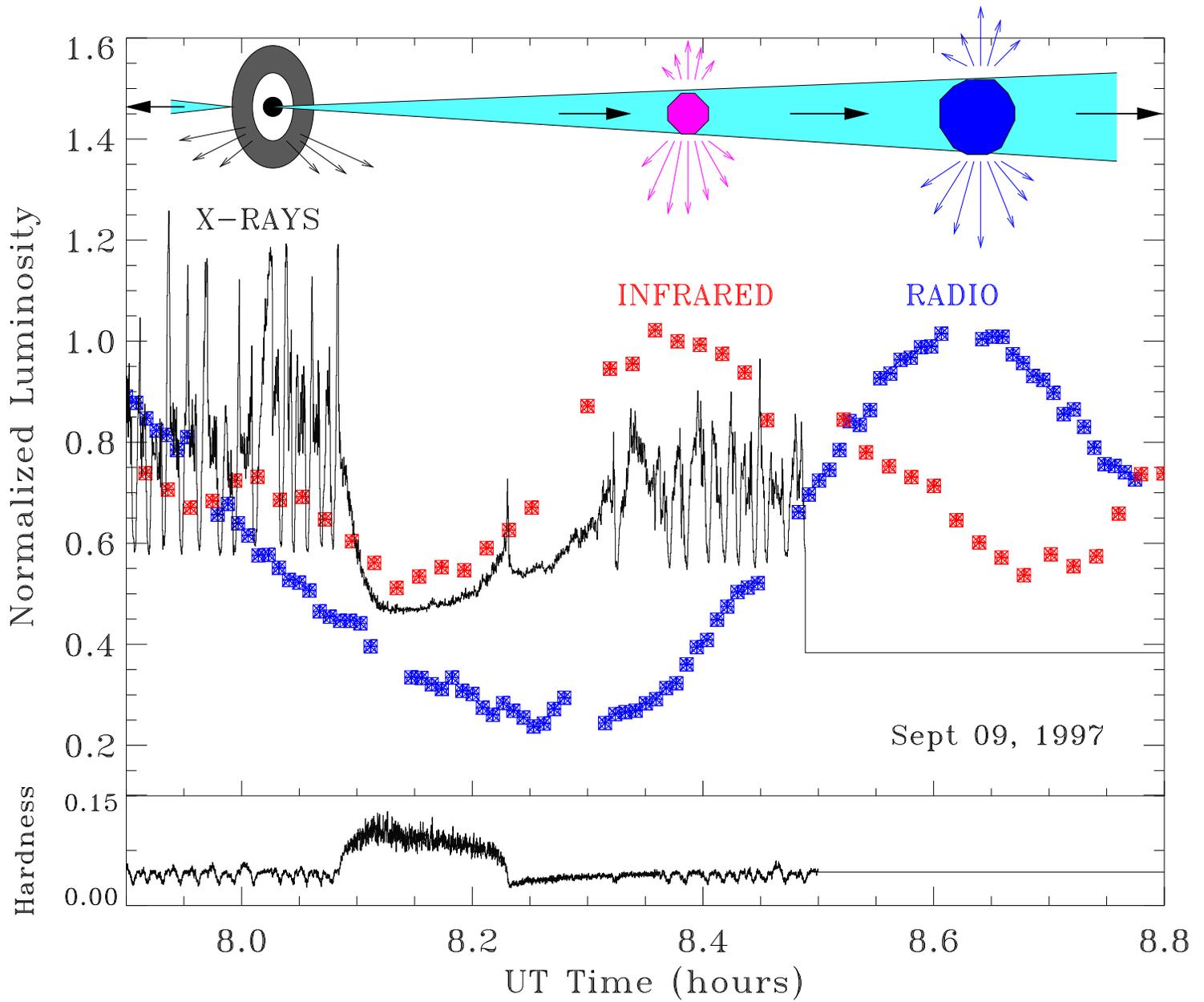,width=1.2\textwidth,angle=270}}
\caption{Contemporaneous X-ray (solid line) radio (dark squares), infrared
  (grey squares) light curves (top panel) and X-ray hardness ratio 
  (bottom panel) for \grs\ on 9 Sep 1997 [45]. 
  The sudden drop in X-ray intensity 
  combined with the X-ray spectral change suggest an emptying of the inner
  accretion disk [25, 26]. The temporal displacement of the
  infrared and radio curves is consistent with being due to synchrotron
  emission from an adiabatically expanding plasmon. (from [45]) }
\label{irrad1915}
\vspace{-0.1cm}
\end{figure}

\section*{5. disk instabilities and Jet ejection}

The variability of the X-ray intensity and spectrum can be rather
complex.
In \grs\ (Fig. 1) periodically repeating stalls, short outbursts, rapid
oscillations
with large amplitude have been observed occasionally while at other
times
the emission is practically constant [25]. In most cases, intensity
variations
are coupled to spectral variability. While not all of the variability
is understood yet, some patters are most probably due to instabilities
in the
accretion disk. During the repeating stalls (Fig. 1) the spectrum
softens dramatically
(factor of $\sim$2 in effective temperature) which has been
interpreted as 
the vanishing of the inner part of the accretion disk [25, 26]
(note, however, that this interpretation has been called into question
because temperature $\equiv$ inner-disk-radius variations are imitated
by a change of the hardening factor $T_{\rm col}/T_{\rm eff}$ when the
accretion rate changes [43]).
After a few minutes the disk hole is refilled, and the disk temperature
and X-ray intensity goes back to its normal, high value. The whole cycle 
of emptying and replenishment of the inner disk is governed by only one 
parameter, namely the radius of the disk ring which empties. The larger
the radius, the deeper the stall, the cooler the measured temperature of the
remaining disk and the longer the time to refill the hole [26], just as 
expected for a radiation-pressure dominated disk.

How, if at all, are these instabilities related to a transient radio emission
and jet ejection? It has been known for several years that transient
radio emission is associated to X-ray state transitions. Direct radio
imaging, discrete peaks in the radio light curves as well as rapidly
evolving radio spectra (to optically thin) suggest that the transient radio
emission is related to discrete ejections.
The most dramatic observation showing the relation of transient radio
emission to the disk behaviour
was made for \grs\ in September 1997 (Fig. 3) when radio and infrared 
oscillations have been found to follow large-amplitude X-ray variations
[45] similar to those described above (Fig. 1). These radio/infrared 
oscillations
have been interpreted as synchrotron emission from repeated small ejections
suffering adiabatic expansion losses [44, 45].
The time delay between the infrared
and radio maximum is consistent with what one expects from an adiabatically
expanding plasma cloud. Together with the above described accretion
disk instability cycle this strongly suggests that the inner part of
the accretion disk disappears as X-ray radiating source, and possibly
is accelerated and ejected away from the system.
First-order estimates suggest that up to $\approx$10\% of the mass
involved in the disk instability may get ejected [26, 44, 45].

Many details of the disk-jet coupling are still unknown.
The expected time delay between the infrared flare and the ejection 
(X-ray event) is only 10$^{-3}$ sec, and thus it is unclear which
X-ray feature would relate to the ejection. Alternatively, one could
interprete this observed time delay 
as the duration of a continuous ejection event [45].
Then, the ejection time scale would be related to the viscous time
scale of the disk rather than the dynamical one. Is indeed a fraction
of the inner disk ejected while falling onto the black hole, or does the
innermost disk gets geometrically thick during the instability?
What is the role of magnetic fields?


\section*{6. Differences and similarities between microquasars and quasars}

There are two distinct differences between microquasars and quasars:
\begin{itemize}
\item\vspace{-0.25cm} Jets of quasars are oriented within a few degrees
towards us while those of microquasars are mostly perpendicular. This
is a pure observational bias: statistically one expects the same
number of objects with angles between 0\grad--60\grad\ and
between 60\grad--90\grad. For quasars, however, the jet emission is
intrinsically too faint to be visible without the Doppler boosting
at small angles (factor $>$10$^4$ flux enhancement). This is also the 
reason why quasar jets are one-sided
only while in microquasars they are two-sided.
It is interesting to speculate how a microquasar would look like if
its beam is directed towards us! 
\item\vspace{-0.22cm} The masses of the accreting objects in microquasars are
a factor of 10$^{6-9}$ smaller than those of quasars. Most of the 
basic observable  parameters scale with this mass. Thus, microquasars
provide the fortunate circumstance that the timescale of accretion and
jet ejection is much better adapted to the human lifetime than that of
quasars.
\vspace{-0.2cm}
\end{itemize}

The surprising finding that the two microquasars \gro\ and \grs\ may
contain maximally rotating black holes actually fits nicely into our
picture of quasars.
Black hole rotation has long been considered as a way to explain the 
radio-loud vs. radio-quiet dichotomy in AGN [46].
While the impact of the black hole spin on jet ejection has still to be
clarified, this relation provides one of the many examples that
the similarities between microquasars and quasars go beyond their names.


\section*{7. Outlook}

For the near future one can expect further substantial progress
because the good observability of the ``right'' timescales of the 
jet-disk coupling (seconds to minutes as compared to many years in
quasars) combined with a series of new generation instrumentation at
X-ray (Chandra, XMM, Astro-E) and Earth-bound (8--10 m class
telescopes) optical wavelength opens new perspectives to attack some
of the major unsolved questions:
\begin{itemize}
\item\vspace{-0.2cm} 
If it is possible to identify emission lines originating in the jets
and to measure their Doppler motion, a new and independent method for
distance determination would be available [5].
\item\vspace{-0.2cm} 
Our present understanding allows a much better prediction of the
disk-jet interaction which in turn will improve the ability to obtain
much improved simultaneous coverage of the jet ejection events.
\item\vspace{-0.2cm} 
X-ray observations with high spectral resolution will provide
new insight into the dynamics of matter flow and emission processes
in the strong gravitational field near black holes. One may expect
that the spin and the mass of black holes could be determined thus
providing the basis for the understanding of the energy source of jet
ejection and acceleration to relativistic speeds.
\item\vspace{-0.2cm} 
The new correlations found between QPO properties and spectral
characteristics will eventually lead to a better theoretical 
understanding of the origin of QPOs which in turn promises to measure
the spin and the mass of black holes and the intimate connection
between accretion disk instabilities and jet ejection.
\end{itemize}

\bigskip
\noindent {\it Acknowledgements:
JG is supported by the German Bun\-des\-mi\-ni\-sterium f\"ur Bildung,
Wissenschaft, Forschung und Technologie
(BMBF/DLR) under contract Nos. 50 QQ 9602 3, and
expresses gratitude for support from the organizers and DFG grants
KON 1973/1999 and GR 1350/7-1.
}

\end{document}